\documentclass[paper]{ptephy_v1} 

\usepackage{bm}

\usepackage{color}
\usepackage{ulem}

\newcommand{\nc}{\newcommand}           
\nc{\vc}[1]     {\mbox{\boldmath $#1$}} 
\nc{\wtil}      {\widetilde}            
\nc{\bras}[1]   {\langle#1|}            
\nc{\kets}[1]   {|#1\rangle}            
\nc{\bra}       {\langle}            
\nc{\ket}       {\rangle}            
\nc{\hO}        {O}           
\nc{\HO}        {\widehat{O}}   

\begin{document}

\title{Soft dipole resonance in neutron-rich $^8$He}

\author[1,2]{Takayuki Myo}
\affil[1]{General Education, Faculty of Engineering, Osaka Institute of Technology, Osaka 535-8585, Japan}
\affil[2]{Research Center for Nuclear Physics (RCNP), Osaka University, Ibaraki, Osaka 567-0047, Japan}

\author[3]{Myagmarjav Odsuren}
\affil[3]{School of Engineering and Applied Sciences, Nuclear Research Centre, National University of Mongolia, Ulaanbaatar 210646, Mongolia}
  
\author[4]{Kiyoshi Kat\=o}
\affil[4]{Nuclear Reaction Data Centre, Faculty of Science, Hokkaido University, Sapporo 060-0810, Japan}


\begin{abstract}
  In neutron-rich $^8$He, we study the soft dipole resonance, which is regarded as a dipole oscillation of four valence neutrons against the $^4$He core,
  and its effect on the low-energy electric dipole strength with a $^4$He+$n$+$n$+$n$+$n$ five-body cluster model.
  This work is an extended study of an earlier letter [T. Myo and K. Kat\=o, Phys. Rev. C{\bf 106}  L021302 (2022)].
  The five-body unbound $1^-$ states of $^8$He are obtained with the complex energy eigenvalues by using the complex scaling method
  and the dipole strength is calculated in terms of the complex-scaled Green's function.
  Two kinds of the dominant excitation modes are confirmed in the dipole strength below 20 MeV of the excitation energy.
  The strengths below 10 MeV are exhausted by the $^7$He+$n$ channel, which sequentially decays to $^6$He+$n$+$n$.
  Above 10 MeV, the strengths arise from the soft dipole mode of four neutrons ($4n$) oscillating against the $^4$He core.
  We further explore the possibility of the soft dipole resonance for this state by carefully searching for the resonance pole 
  and finally predict the corresponding resonance with the excitation energy of 14 MeV and the decay width of 21 MeV.
  The soft dipole resonance exhausts about half of the dipole strength in the relative motion between the $^4$He core and $4n$.
\end{abstract}

\subjectindex{D10, D11, D13}

\maketitle 

\section{Introduction}

Unstable nuclei have shown exotic nuclear structures with the development of radioactive beam experiments.
Neutron halo structure is observed in some of the drip-line nuclei, such as $^6$He and $^{11}$Li \cite{tanihata85,tanihata13}.
In unstable nuclei, excess nucleons often form the weakly bound state with respect to the stable core nucleus and they are excited to the unbound states beyond the particle thresholds with small excitation energy.
Hence, the structures of resonances and the responses of continuum states become the important research targets of unstable nuclear physics.
A soft dipole mode including the possibility of resonance has been proposed as a new collective motion with dipole oscillation of excess neutrons against the core nucleus \cite{hansen87,ikeda92}.
Here we define that the core nucleus is kept in the ground state in the soft dipole mode in comparison with the giant dipole resonances, in which the core nucleus is excited.
The Coulomb breakup reaction is useful to investigate the dipole response of unstable nuclei experimentally,
and can bring exotic excitations due to the presence of excess neutrons \cite{nakamura94,aumann99,nakamura06}.

In neutron-rich He isotopes, only two bound states are observed in $^6$He and $^8$He.
The $^8$He nucleus has a large neutron-proton ratio of 3 and consists of the $^4$He core nucleus and four valence neutrons,
which are bound with small separation energy of 3.1 MeV in the ground state \cite{tilley04}.  
For the excitations of $^8$He, many experiments have been reported \cite{korsheninnikov93,iwata00,meister02,chulkov05,mueller07,golovkov09,holl21},
and recently, the dipole excitation is examined theoretically \cite{nonaiti22,piekarewicz22}.

In $^8$He, the lowest particle-threshold is the $^6$He+$n$+$n$ three-body channel with the excitation energy of 2.1 MeV,
and the next channel is the $^7$He+$n$ two-body one with the excitation energy of 2.6 MeV, where the ground state of $^7$He is a resonant state of $^6$He+$n$.
Experiments report the candidates of the excited states of $^8$He, which are located above the $^4$He+$n$+$n$+$n$+$n$ five-body channel \cite{golovkov09,holl21}.
From these facts, the excitations of $^8$He lead to the breakups into many-body channels of $^7$He+$n$, $^6$He+$n$+$n$, $^5$He+$n$+$n$+$n$, and $^4$He+$n$+$n$+$n$+$n$.

So far, we have performed the analysis of the neutron-rich He isotopes and their mirror proton-rich unbound nuclei in the $^4$He+$N+N+N+N$ five-body cluster model, where $^4$He is treated as an inert core \cite{myo10,myo12,myo14b,myo21,myo22}.
We solve the motion of multi valence-nucleons around the $^4$He core in the cluster orbital shell model (COSM) \cite{suzuki88,masui06,myo07a,masui12}.
In the COSM, the threshold energies of the particle emissions can be reproduced, namely, the subsystem energies of $^8$He.
This is an important property in the proper description of the multineutron emissions.
Under this condition, we describe many-body resonances using the complex scaling method (CSM) \cite{ho83,moiseyev98,aoyama06,moiseyev11,myo14a,myo20} imposing the correct boundary conditions for decaying states.
In the CSM, one solves the eigenvalue problem of the complex-scaled Hamiltonian using $L^2$ basis functions and obtains the resonances explicitly with complex energy eigenvalues.
Structures of resonances have been successfully investigated using the CSM not only for eigenenergies but also for the strength function applying the Green's function \cite{myo14a,myo20,myo98,myo01,suzuki05,odsuren15}. 
For Coulomb breakup reactions of halo nuclei $^6$He and $^{11}$Li, we have successfully investigated the three-body breakup cross sections \cite{myo14a,myo01,kikuchi10,kikuchi13}.
In particular, we can decompose the strength function into the various channels of the direct three-body breakup and the sequential breakup via the subsystems.

In our previous work \cite{myo21}, we predicted four resonances of $^8$He with positive parity in the low excitation energy and the lowest resonance is the $2^+$ state,
whose energy and decay width are consistent with the recent experimental data \cite{holl21}.
In our recent letter \cite{myo22}, we predicted the soft dipole mode of $^8$He ($1^-$) in the electric dipole strength,
which makes a mild peak at the excitation energy of 13 MeV. This mode shows a collective nature of four neutrons oscillating against the $^4$He core.
In this paper, we report the detailed and extended analysis of the electric dipole strength of $^8$He in the low excitation-energy according to the recent letter.
We calculate the $1^-$ states of $^8$He in the COSM and CSM and evaluate the dipole strength function using the Green's function.
For the soft dipole mode, we further explore the possibility of the isolated resonance by searching for the resonance pole carefully in the complex scaling.
We do not discuss the giant resonances in this study, because the configuration of $^4$He is fixed in the five-body cluster model.
The present analysis is useful for the Coulomb breakup experiments of $^8$He \cite{lehr22}.

In section~\ref{sec:method}, we explain the framework of the COSM with the CSM. 
In section~\ref{sec:result}, we discuss the results of the electric dipole strength of $^8$He and the possibility of soft dipole resonance.
In section~\ref{sec:summary}, we give a summary.

\section{Method}\label{sec:method}

\subsection{Cluster orbital shell model}

\begin{figure}[t]
\centering
\includegraphics[width=4.5cm,clip]{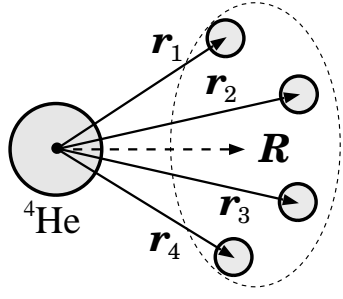}
\caption{Coordinate system of the $^4$He+$n$+$n$+$n$+$n$ in the COSM.
The vector $\vc{R}$ is a relative coordinate from the center-of-mass of $^4$He to the center-of-mass of four neutrons.}
\label{fig:COSM}
\end{figure}

We explain the five-body cluster model of $^8$He with the $^4$He+$n$+$n$+$n$+$n$ COSM.
The relative coordinates of four neutrons are $\{\vc{r}_i\}$ with $i=1,\ldots,4$ as shown in Fig.~\ref{fig:COSM}.
The Hamiltonian is the same as used in the previous studies~\cite{myo10,myo12,myo14b,myo21,myo22}:
\begin{eqnarray}
	H
&=&	t_\alpha+ \sum_{i=1}^{A_v} t_i - t_G + \sum_{i=1}^{A_v} v^{\alpha N}_i + \sum_{i<j}^{A_v} v^{NN}_{ij}
    \\
&=&	\sum_{i=1}^{A_v} \left( \frac{\vc{p}^2_i}{2\mu} + v^{\alpha N}_i \right) + \sum_{i<j}^{A_v} \left( \frac{\vc{p}_i\cdot \vc{p}_j}{A_\alpha m} + v^{NN}_{ij} \right) ,
    \label{eq:Ham}
\end{eqnarray}
where $A_v$ and $A_\alpha$ are the number of valence neutrons and a mass number of $^4$He, respectively.
The total mass number is $A=A_v+A_\alpha$, where $(A_v, A_\alpha)=(4,4)$ for $^8$He.
The kinetic energy operators $t_\alpha$, $t_i$, and $t_G$ are those of $^4$He, one neutron, and the center-of-mass part, respectively.
The operator $\vc{p}_i$ is the relative momentum between $^4$He and a neutron.
The $^4$He--nucleon interaction $v^{\alpha N}$ is given by the microscopic Kanada-Kaneko-Nagata-Nomoto potential \cite{aoyama06,kanada79}.
For nucleon-nucleon interaction $v^{NN}$, we use the Minnesota central potential \cite{tang78}.

In the COSM, the $^8$He wave function $\Psi^J$ with spin $J$ is expanded in the linear combination form of the COSM configuration $\Psi^J_c$ as
\begin{eqnarray}
    \Psi^J
&=& \sum_c C^J_c \Psi^J_c,
    \label{WF0}
    \\
    \Psi^J_c
 &=& {\cal A}' \left\{ \Phi(^4{\rm He}), \Phi^J_c \right\},
    \\
    \Phi^J_c
&=& {\cal A} \left[ \left[ \phi_{p_1}(\vc{r}_1), \phi_{p_2}(\vc{r}_2) \right]_{j_{12}},\left[ \phi_{p_3}(\vc{r}_3),\phi_{p_4}(\vc{r}_4) \right]_{j_{34}} \right]_J .
    \label{WF1}
\end{eqnarray}
We adopt the $(0s)^4$ configuration of the harmonic oscillator wave function for the $\Phi(^4{\rm He})$ of $^4$He.
The range parameter of the $0s$ orbit is 1.4 fm to reproduce the charge radius of $^4$He.
For valence neutrons, the single-particle wave function $\phi_p(\vc{r})$ is a function of relative coordinate $\vc{r}$ and
has a quantum number $p$, the set of $\{n,\ell,j\}$ in the $jj$ coupling scheme.
The label $n$ is for the different radial component explained later and $\ell$ is the orbital angular momentum coupled with a neutron spin as $j=[\ell,1/2]$.
The spins of $j_{12}$ and $j_{34}$ are for the neutron pairs.
The operators for the antisymmetrization ${\cal A'}$ and ${\cal A}$ are between $^4$He and a valence neutron and between valence neutrons, respectively.
The former is treated with the orthogonality condition model \cite{saito69}, where the relative $0s$ orbit is removed in the $\phi_p$.
The label $c$ in Eq.~(\ref{WF0}) means the set of $\{p_1,p_2,p_3,p_4,j_{12},j_{34}\}$.
We superpose the available configurations $\Psi^J_c$ with the amplitude of $C_c^J$ in Eq.~(\ref{WF0}).
The Hamiltonian matrix elements are calculated using the COSM basis states in the analytical form
and one solves the eigenvalue problem of the Hamiltonian:
\begin{eqnarray}
  \sum_{c'}\langle \Psi^J_c |H|  \Psi^J_{c'} \rangle\, C^J_{c'} &=& E^J C_{c}^J  ,
  \label{eq:eigen}
  \\
  \sum_c (C^J_{c})^2&=&1.
\end{eqnarray}
We obtain the amplitudes $\{C^J_c\}$ and the energy eigenvalues $E^J$ of $^8$He
measured from the five-body threshold energy of $^4$He+$n$+$n$+$n$+$n$.

We expand the radial part of $\phi_p(\vc{r})$ with a finite number of Gaussian functions $u(\vc{r},b)$ for each single-particle state:
\begin{eqnarray}
    \phi_p(\vc{r})
&=& \sum_{q=1}^{N_{\ell j}} d^q_p\ u_{\ell j}(\vc{r},b_{\ell j}^q)\, ,
    \label{spo}
    \\
    u_{\ell j}(\vc{r},b_{\ell j}^q)
&=& r^{\ell} e^{-(r/b_{\ell j}^q)^2/2}\, [Y_{\ell}(\hat{\vc{r}}),\chi^\sigma_{1/2}]_{j}\, ,
    \label{Gauss}
	\\
    \langle \phi_p | \phi_{p'} \rangle 
&=& \delta_{p,p'}
~=~ \delta_{n,n'}\, \delta_{\ell,\ell'}\, \delta_{j,j'}.
    \label{Gauss2}
\end{eqnarray}
The label $q$ is for the Gaussian range parameter $b_{\ell j}^q$ with $q=1,\ldots, N_{\ell j}$,
where $N_{\ell j}$ is a basis number.
The parameters $\{b_{\ell j}^q\}$ are given in the geometric progression \cite{hiyama03}.
The coefficients $\{d^q_p\}$ in Eq.~(\ref{spo}) are determined using the orthogonal condition of the basis states $\{\phi_p\}$ in Eq.~(\ref{Gauss2}).
The number $N_{\ell j}$ is determined to get the convergence of the solutions and we use $N_{\ell j}=12$ at most
with the range of $b_{\ell j}^q$ from 0.3 fm to around 40 fm.
The multineutron configuration $\Phi^J_c$ in Eq.~(\ref{WF1}) is expanded using the products of the single-particle basis states $\phi_p$.

For the single-particle states $\phi_p$, we include the orbital angular momenta $\ell\le 2$ for the ground state of $^8$He.
This condition gives the two-neutron separation energy of $^6$He($0^+$) with the accuracy of 0.3 MeV
in comparison with the calculation including a large $\ell$.
For $1^-$ states of $^8$He, we increase the configurations including the $\ell=3$ states for the dipole transition.
We use the 173.7 MeV of the repulsive strength of the Minnesota potential $v^{NN}$ instead of the original 200 MeV
to reproduce 0.975 MeV of the two-neutron separation energy of $^6$He.
This condition is the same as used in the previous works \cite{myo10,myo12,myo21,myo22}
and gives a nice reproduction of the energy levels of He isotopes and their mirror nuclei.

\subsection{Electric dipole transition}

We investigate the electric dipole transition from the ground state of $^8$He.
In the present five-body COSM, we fix the configuration of $^4$He, 
and consider the dipole transition induced by the motion of valence neutrons around $^4$He.
We introduce the relative coordinate $\vc{R}$ between $^4$He and the center-of-mass of four valence neutrons as shown in Fig. \ref{fig:COSM}.
The operator of the electric dipole ($E1$) transition is expressed in terms of  $\vc{R}$ with $R=|\vc{R}|$ as the recoil effect of $^4$He from the center-of-mass, 
\begin{eqnarray}
  \hO_{E1,\mu}&=& -\frac{A_v}{A} eZ_\alpha R\, Y_{1\mu}(\widehat{\vc{R}}), 
  \\
  \vc{R}&=& \frac{1}{A_v}\sum_{i=1}^{A_v}\vc{r}_i,
\end{eqnarray}
where $Z_\alpha$ is a proton number of $^4$He.
We consider the completeness relation of the $1^-$ states of $^8$He in the dipole transition,
which brings the non-energy weighted sum-rule value for the cluster model, denoted as $B_{\rm c}(E1)$
and expressed using the mean squared distance between $^4$He and center-of-mass of four valence neutrons in the ground state,
\begin{eqnarray}
  B_{\rm c}(E1)&=& \sum_\mu \langle \wtil{\Phi}_{0^+} | \hO^\dagger_{E1,\mu} \hO_{E1,\mu} |\Phi_{0^+}\rangle
~=~ \frac{3e^2}{4\pi}\cdot \left(\frac{A_vZ_\alpha}{A}\right)^2 \langle \vc{R}^2 \rangle.
\label{eq:SRV}
\end{eqnarray}
The value of $B_{\rm c}(E1)$ is useful to evaluate the amount of dipole strength obtained in the calculation. 
For comparison, taking the $A$-nucleon dipole excitations, the ordinary non-energy weighted sum-rule value is
given as $B(E1)=3e^2/(4\pi)\cdot Z_\alpha \langle r^2_p \rangle$
where $\langle r^2_p \rangle$ is a mean squared proton radius of the ground state.
In this study, the excitation of $^4$He is not treated and
this effect is expected to contribute to the dipole strength of $^8$He above the excitation energy of around 20 MeV.

\subsection{Complex scaling method}

We describe resonances and continuum states in the many-body systems with the CSM \cite{ho83,moiseyev98,moiseyev11,aoyama06,myo14a,myo20}.
In this study, resonances are the Gamow states satisfying the outgoing boundary condition with the complex eigenenergies,
and the continuum states are orthogonal to these resonances.
In the CSM, the coordinates $\{\vc{r}_j\}$ in Fig. \ref{fig:COSM} are transformed with a scaling angle $\theta$ as
\begin{eqnarray}
  \vc{r}_j \to \vc{r}_j\, e^{i\theta},\qquad
  \vc{p}_j \to \vc{p}_j\, e^{-i\theta},
  \label{CSM}
\end{eqnarray}
where $\vc{p}_i$ is the conjugate momentum of $\vc{r}_i$ and is used in the Hamiltonian Eq.~(\ref{eq:Ham}).
The complex-scaled Schr\"odinger equation is written with the complex-scaled Hamiltonian $H_\theta$ as 
\begin{eqnarray}
	H_\theta   \Psi^J_\theta
&=&     E^J_\theta \Psi^J_\theta ,
	\label{eq:eigen2}
        \\
    \Psi^J_\theta
&=& \sum_c C^J_{c,\theta} \Psi^J_c,
    \label{WF_CSM}
    \\
1&=& \sum_{c} \big(C^J_{c,\theta}\big)^2.
    \label{norm_CSM}
\end{eqnarray}
We solve the eigenvalue problem of Eq.~(\ref{eq:eigen2}) and
obtain the complex-scaled wave function $\Psi^J_\theta$ in Eq.~(\ref{WF_CSM}),
where the coefficients $C_{c,\theta}^J$ are $\theta$-dependent complex numbers.
We obtain the energy eigenvalues $E^J_\theta$ on a complex energy plane according to the so-called ABC theorem proved by Aguilar, Balslev, and Combes \cite{ABC}.

This theorem indicates that the complex scaling transforms the divergent outgoing resonant wave into a damping form.
As a result, the asymptotic boundary condition of resonances in the CSM becomes the same as bound states \cite{aoyama06,myo14a}.
In the CSM, every Riemann branch cut is rotated down by $2\theta$ in the complex energy plane,
according to the momentum transformation in Eq.~(\ref{CSM}).
The branch cuts start from the threshold energies of particle emissions
and the continuum states are obtained on the corresponding $2\theta$ lines.
In the ABC theorem, the energies of bound and resonant states are independent of $\theta$
and the resonance energy eigenvalue is given as $E_r-i\Gamma/2$
with the resonance energy $E_r$ measured from the threshold energy and the decay width $\Gamma$.

In the CSM, one can expand the resonance wave functions with the $L^2$ basis functions
because of the damping boundary condition, being normalized in the condition in Eq.~(\ref{norm_CSM}).
One does not use the Hermitian product according to the bi-orthogonal property of the adjoint states \cite{ho83,moiseyev98,berggren68}.

We calculate the strength function of the electric dipole transition into the unbound states of $^8$He.
For this purpose, we consider the extended completeness relation (ECR) of $^8$He consisting of bound, resonant, and continuum states
with the complex-scaled eigenstates $\Psi^J_\theta$ in Eq.~(\ref{eq:eigen}) \cite{myo01,myo10,berggren68}.
When one adopts a large value of $\theta$, five-body components of $^8$He are classified into several categories,
which construct the five-body ECR of $^8$He as
\begin{eqnarray}
	{\bf 1}
&=&	\sum_{~\nu} \kets{\Psi^J_{\theta,\nu}}\bras{\wtil{\Psi}^J_{\theta,\nu}}
	\nonumber
	\\
&=&	\{\mbox{bound state of $^8${He}}\}
	\nonumber
	\\
&+&	\{\mbox{resonances of $^8${He}}\}
	\nonumber
	\\
&+&	\{\mbox{two-body continuum states of $^7${He}$^{(*)}+n$}\}
	\nonumber
	\\
&+&	\{\mbox{three-body continuum states of $^6${He}$^{(*)}+n+n$}\}
	\nonumber
	\\
&+&	\{\mbox{four-body continuum states of $^5${He}$^{(*)}+n+n+n$}\}
	\nonumber
	\\
&+&	\{\mbox{five-body continuum states of $^4${He} $+n+n+n+n$}\} ,
	\label{eq:ECR}
\end{eqnarray}
where $\{ \Psi^J_{\theta,\nu},\wtil{\Psi}^J_{\theta,\nu} \}$ form biorthogonal bases with the state index $\nu$.

We explain the general procedure to calculate the strength function using the ECR, where we omit the spin notation.
One can define the complex-scaled Green's function ${\cal G}^\theta(E)$ as a function of the energy $E$:
\begin{eqnarray}
	{\cal G}^\theta(E)
&=&	\frac{ {\bf 1} }{ E-H_\theta }
~=~	\sum_{~\nu}
	\frac{|\Psi^\theta_\nu\rangle \langle \wtil{\Psi}^\theta_\nu|}{E-E_\nu^\theta} .
	\label{eq:green1}
\end{eqnarray}
The strength function $S_\lambda(E)$ of the transition operator $\hO_\lambda$ with rank $\lambda$
is represented by using the Green's function without the complex scaling as
\begin{eqnarray}
	S_\lambda(E) 
&=&	\sum_{~\nu}
	\bras{\wtil{\Psi}_0}\hO_\lambda^\dagger\kets{\Psi_\nu}\bras{\wtil{\Psi}_\nu}\hO_\lambda\kets{\Psi_0}\,
	\delta(E-E_\nu)
	\label{eq:strength_org}
	\\
&=&	-\frac1{\pi}\ {\rm Im}\left\{ \langle \wtil{\Psi}_0 | \hO^\dagger_\lambda {\cal G}(E) \hO_\lambda | \Psi_0 \rangle \right\}, 
	\label{eq:strength1}
\end{eqnarray}
where $\Psi_0$ is the initial state.
We apply the complex scaling to the strength function and use ${\cal G}^\theta(E)$ in Eq.~(\ref{eq:green1}).
\begin{eqnarray}
	S_\lambda(E)
&=&     \sum_{~\nu} S_{\lambda,\nu}(E) ,
	\label{eq:strength2}
        \\
        S_{\lambda,\nu}(E)
&=&     -\frac1{\pi}\ {\rm Im}\left\{  \frac{
	\bras{\wtil{\Psi}_0^\theta}  (\hO_\lambda^\dagger)^\theta \kets{\Psi_\nu^\theta}
	\bras{\wtil{\Psi}_\nu^\theta} \hO_\lambda^\theta          \kets{\Psi_0^\theta}
        }{E-E_\nu^\theta}
        \right\}. 
	\label{eq:strength3}
\end{eqnarray}
One can extract the contributions of the state $\nu$, $S_{\lambda,\nu}(E)$, in the total strength $S_\lambda(E)$
and classify $S_\lambda(E)$ in terms of the ECR in Eq. (\ref{eq:ECR}).
It is noted that the functions $S_{\lambda}(E)$ and $S_{\lambda,\nu}(E)$ are independent of $\theta$ \cite{myo01,myo10,odsuren15}.
This is because the state $\nu$ is uniquely classified in the ECR in Eq.~(\ref{eq:ECR}) and then $S_{\lambda,\nu}(E)$ is also uniquely obtained.
The complex-scaled Green's function has been widely used in calculations of the scattering amplitudes
\cite{kikuchi10,myo14a,kruppa07,dote13}.

We mention the properties of the strength function $S_{\lambda,\nu}(E)$.
The total strength function $S_{\lambda}(E)$ is the observable and positive definite.
On the other hand, the component $S_{\lambda,\nu}(E)$ is not imposed to keep the positive definite at all energies,
because $S_{\lambda,\nu}(E)$ is non-observable, similar to the complex energies of resonances.
This property means that $S_{\lambda,\nu}(E)$ can show negative values.
We have generally discussed this point of the strength function in the CSM in Refs. \cite{myo98,myo01}.

In this study, many-body continuum states are expanded with the Gaussian functions in Eq.(\ref{Gauss}) similar to the resonances,
and then the states are discretized.
The reliability of the discretization of the unbound states in the CSM has been shown
in terms of the continuum level density \cite{suzuki05,kruppa07,odsuren14,odsuren21}.

\section{Results}\label{sec:result}

\subsection{Energy spectra of He isotopes}

\begin{figure}[b]
\centering
\includegraphics[width=10.5cm,clip]{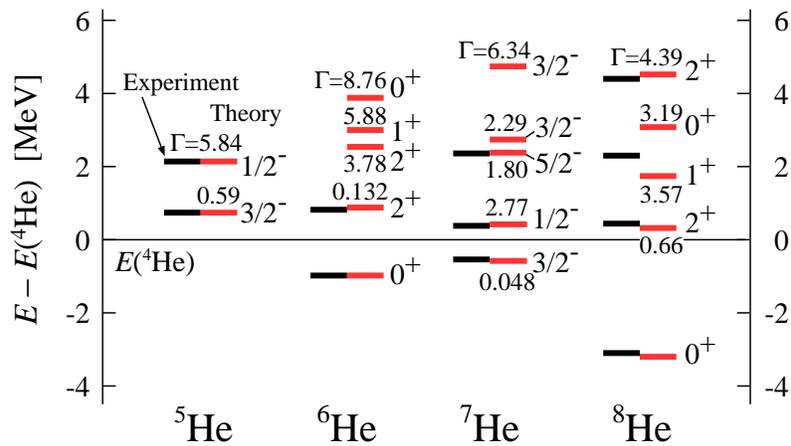}
\caption{Energy levels of $^{4-8}$He measured from the $^4$He energy \cite{myo21}. Units are in MeV.
  Red (right) and black (left) lines are the values of theory and experiments, respectively.
  Small numbers indicate the decay widths $\Gamma$ of resonances.
  For $^8$He, the experimental data are taken from Refs. \cite{golovkov09,holl21}.}
\label{fig:He}
\end{figure}

We start to discuss the results of the energy spectra of He isotopes to show the validity of the COSM with the complex scaling,
the main parts of which are already reported in the previous work \cite{myo21}.
The energy spectra are useful to confirm the threshold energies of the multineutron emissions in the dipole transition of $^8$He.
The systematic behavior of the energy levels of $^{4-8}$He is shown in Fig. \ref{fig:He}.
For $^8$He, we obtain five states with positive parity; only the ground state is a bound state and the others are resonances.

For the ground $0^+$ state of $^8$He, the relative energy measured from the $^4$He+$n$+$n$+$n$+$n$ threshold is 3.22 MeV and close to the experimental value of 3.11 MeV.
For the $2^+_1$ resonance of $^8$He, we obtain the relative energy of 0.32 MeV and the decay width $\Gamma$ of 0.66 MeV.
These values are consistent with the recent experimental report of the corresponding energy 0.43(6) MeV and decay width $\Gamma$=0.89(11) MeV,
which are derived from the energy dependence of the cross section in the proton inelastic scattering \cite{holl21}.
The dominant configurations of four neutrons in the energy levels are shown in the previous analysis \cite{myo21}.

\begin{table}[t]  
  \caption{Radial properties of the ground state of $^8$He in units of fm for matter, proton, neutron, and charge parts.
Comparison with the experiments; a\cite{tanihata92}, b\cite{alkazov97}, c\cite{kiselev05}, d\cite{mueller07}, e\cite{brodeur12}, and f\cite{krauth21}, are shown.
Mean distances between $^4$He and one valence neutron $\alpha\mbox{-}n$ and between $^4$He and four valence neutrons $\alpha\mbox{-}4n$, and the radius of four valence neutrons $4n$, are shown together in fm.
The non-energy weighted sum-rule value $B_{\rm c}(E1)$ of the dipole strength in the cluster model is given in units of $e^2$fm$^2$.}
  \label{tab:radius}
  \centering
    \begin{tabular}{c|cc}
      \noalign{\hrule height 0.5pt}
                       & Theory      & Experiments   \\ \hline
      matter           & 2.53        & 2.49(4)$^{\rm a}$~~~2.45(7)$^{\rm b}$~~~2.49(4)$^{\rm c}$ \\
      proton           & 1.81        &     \\
    ~~neutron~~        & 2.73        &     \\
      charge           & 1.92        & 1.929(26)$^{\rm d}$~~1.959(16)$^{\rm e}$~~1.9559(158)$^{\rm f}$  \\ \hline
$\alpha\mbox{-}n$      & 3.56        &     \\ 
$\alpha\mbox{-}4n$     & 2.05        &     \\
$4n$                   & 2.91        &     \\ \hline
$B_{\rm c}(E1)$              & 1.01        &     \\
\noalign{\hrule height 0.5pt}
\end{tabular}
\end{table}

In Table \ref{tab:radius}, we summarize the spatial properties of the ground state of $^8$He.
The matter and charge radii of $^8$He are 2.53 and 1.92 fm, respectively, which agree with the experiments.
The mean distance between $^4$He and the center-of-mass of four valence neutrons, $\alpha\mbox{-}4n$, is obtained as $\sqrt{\langle \vc{R}^2 \rangle}=2.05$ fm.
This distance is responsible for the electric dipole transition from the ground state according to Eq.~(\ref{eq:SRV}).
The non-energy weighted sum-rule value for cluster model $B_{\rm c}(E1)$ is obtained as 1.01 $e^2$fm$^2$ using this distance. 
The radius of four neutrons in $^8$He is also obtained as 2.91 fm.

\subsection{Electric dipole transition}

We investigate the electric dipole transition strength of $^8$He.
Before showing the results with the complex scaling, first we demonstrate the case without the complex scaling, that is,
the scaling angle $\theta=0^\circ$ and the eigenenergies of the $1^-$ states of $^8$He are discretized on the real energy axis in the bound-state approximation.
The transition strengths are calculated for each discretized state and
we adopt Eq.~(\ref{eq:strength_org}) to calculate the dipole strength
\begin{eqnarray}
  B(E1,\nu)=\bras{\wtil{\Psi}_{0^+}}|\hO_{E1}^\dagger|\kets{\Psi_{1^-,\nu}}\bras{\wtil{\Psi}_{1^-,\nu}}|\hO_{E1}|\kets{\Psi_{0^+}}
\end{eqnarray}
using the reduced matrix elements of the state index $\nu$ in the discretized $1^-$ states.

\begin{figure}[t]
  \centering
  \includegraphics[width=8.5cm,clip]{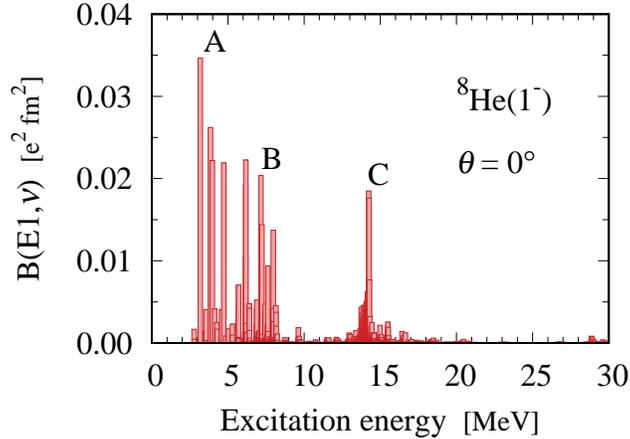}
  \caption{Discretized electric dipole strength $B(E1,\nu)$ of $^8$He with the state index $\nu$ as a function of the excitation energy $E_{x,\nu}$ without complex scaling.
    Units are $e^2$fm$^2$.
    For three states with labels A, B, and C, the transition densities are calculated in section \ref{sec:dns}.
  }
  \label{fig:disc1}
\end{figure}

In Fig.~\ref{fig:disc1}, we show the discretized dipole strength without the complex scaling.
In the distribution, we can confirm the low-energy strengths at around the excitation energies from 3 MeV to 9 MeV
and also at around 14 MeV.
Although the strength distribution is discretized as a function of the energy as shown in Fig.~\ref{fig:disc1},
it becomes continuous by using the Green's function with the complex scaling.
Later, we investigate the properties of the strength using the complex scaling in detail.

\begin{figure}[t]
  \centering
  \includegraphics[width=8.5cm,clip]{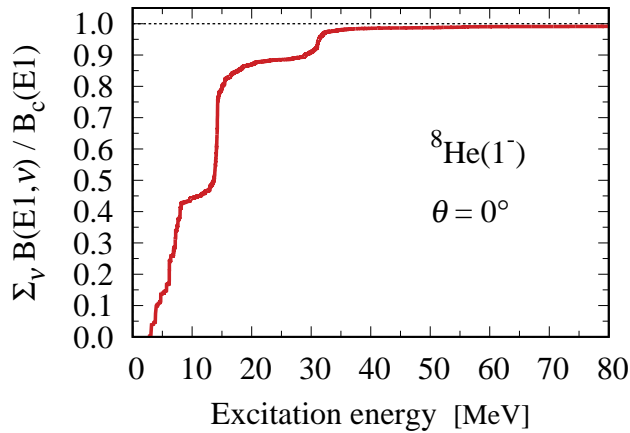}
  \caption{Accumulated electric dipole strength $\sum_\nu B(E1,\nu)$ of $^8$He as a function of the excitation energy without complex scaling.
    The ratio to $B_{\rm c}(E1)$ in Table \ref{tab:radius} is given.}
  \label{fig:disc2}
\end{figure}
  

In Fig.~\ref{fig:disc2}, we show the accumulated dipole transition strength $\sum_\nu B(E1,\nu)$ as a function of the excitation energy of $^8$He.
This is useful to check the completeness relation of the discretized $1^-$ states.
The results are normalized by $B_{\rm c}(E1)$ shown in Table \ref{tab:radius}.
It is found that at 80 MeV, the accumulated strength is converged and the ratio becomes 0.993 with respect to $B_{\rm c}(E1)$.
This means that the $1^-$ states sufficiently construct the completeness relation in the COSM even without the complex scaling.

Next, we perform the calculation of the electric dipole strength using the complex scaling with a finite $\theta$.
One of the advantages of this calculation is that each eigenstate is distinguishable in the categories of various channels of $^8$He explained in Eq.~(\ref{eq:ECR}).
Another advantage is that the complex scaling provides the correct level density even in the finite number of the basis states \cite{suzuki05}.
By using the Green's function with complex scaling in Eq.~(\ref{eq:green1}),
one can obtain the continuous strength as a function of the excitation energy,
being independent of $\theta$, and the resulting distribution can be compared with the experimental data.

\begin{figure}[b]
\centering
\includegraphics[width=9.0cm,clip]{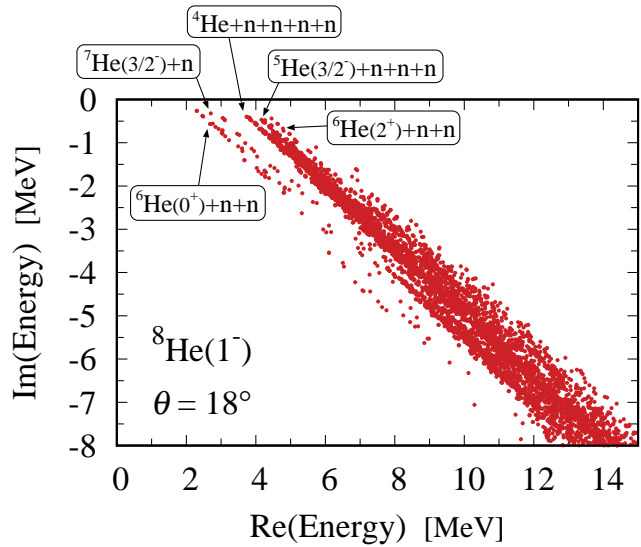}
\caption{
  Energy eigenvalue distribution of $^8$He ($1^-$) with complex scaling in the complex energy plane
  measured from the ground state energy. The scaling angle $\theta$ is 18$^\circ$ \cite{myo22}. Units are in MeV.
  Some of the channels of multineutron emissions are shown.
}
\label{fig:CSM}
\end{figure}

In Fig. \ref{fig:CSM}, we show the complex energy eigenvalues of $^8$He($1^-$) with complex scaling,
measured from the ground state of $^8$He where the scaling angle $\theta=18^\circ$.
The states are obtained along several lines, which have the angle $2\theta$ with respect to the real axis
and are called the $2\theta$ lines.
The $2\theta$ lines are the branch cuts starting from the thresholds, in the case of $^8$He,
of $^6$He($0^+$)+$n$+$n$, $^7$He($3/2^-$)+$n$, $^4$He+$n$+$n$+$n$+$n$, $^5$He($3/2^-$)+$n$+$n$+$n$, and $^6$He($2^+$)+$n$+$n$,
as the excitation energy increases.
These eigenstates represent the discretized continuum states of each channel.
One can extract the contributions of the individual continuum states to the strength function,
which is handled by using the Green's function with complex scaling and ECR \cite{myo98,myo01,myo14a}.

\begin{figure}[t]
\centering
\includegraphics[width=8.5cm,clip]{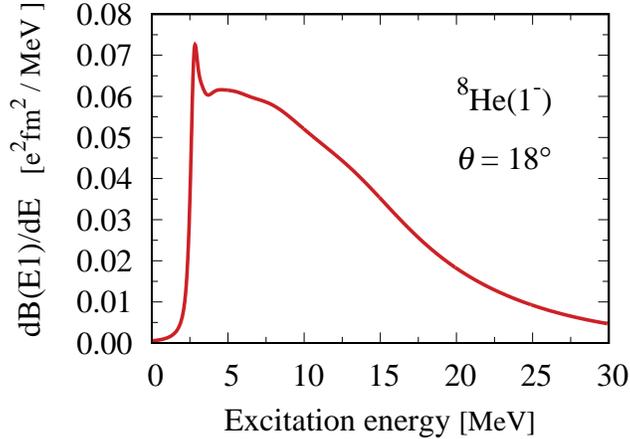}
\caption{Distribution of the electric dipole strength $dB(E1)/dE$ of $^8$He 
  as a function of the excitation energy using the complex scaling with $\theta=18^\circ$ \cite{myo22}. Units are in $e^2$fm$^2$/MeV.}
\label{fig:dist1}
\end{figure}
\begin{figure}[t]
\centering
\includegraphics[width=8.5cm,clip]{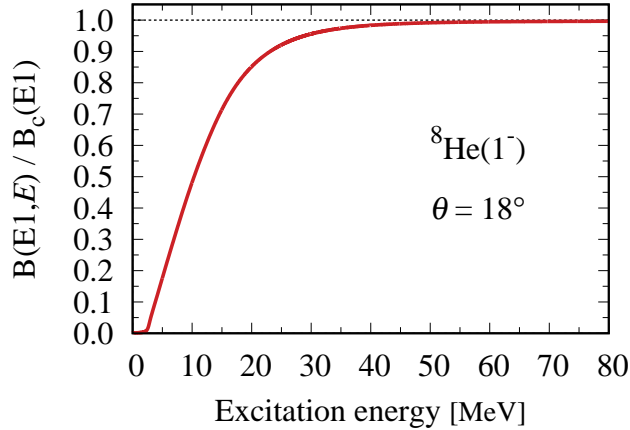}
\caption{Integrated electric dipole strength $B(E1,E)$ of $^8$He using complex scaling up to the excitation energy $E$.
The ratio to $B_{\rm c}(E1)$ in Table \ref{tab:radius} is given.}
\label{fig:dist2}
\end{figure}

In Fig.~\ref{fig:dist1}, we give the electric dipole strength function $dB(E1)/dE$ of $^8$He,
which corresponds to $S_{\lambda=1}(E)$ in Eq.~(\ref{eq:strength2}) as a function of the excitation energy,
where all of the complex-scaled solutions of the $1^-$ states are used.
The line presented in Fig. \ref{fig:dist1}, is obtained by the continuation of discretized strength function shown in Fig. \ref{fig:disc1}.
In the CSM, the strength function obtained with appropriate basis states does not depend on the scaling angle $\theta$.
Some parts of the present analysis have already been reported in the previous letter \cite{myo22}.
It is found that the strength shows a spike at 2.8 MeV and after that, the mild peak is obtained at around 5 MeV.
Above 5 MeV, the strength gradually decreases as the excitation energy increases.

In Fig.~\ref{fig:dist2}, we show the integrated dipole strength up to the excitation energy of $^8$He.
We plot the ratio with respect to $B_{\rm c}(E1)$, similar to the results shown in Fig.~\ref{fig:disc2}.
It is confirmed that at 80 MeV of the excitation energy, the ratio becomes 0.996.
This ensures the completeness relation of the complex-scaled $1^-$ states
and is similar to the case without the complex scaling as shown in Fig. \ref{fig:disc2}.

\begin{figure}[t]
\centering
\includegraphics[width=8.5cm,clip]{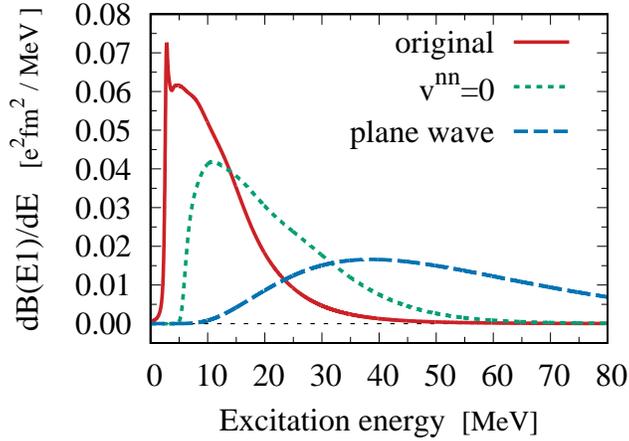}
\caption{Distributions of the electric dipole strengths of $^8$He using different conditions of the final $1^-$ states
  with $\theta=18^\circ$.
  (1) plane waves with a blue dashed line,
  (2) without the interactions between valence neutrons ($v^{NN}=0$) with a green dotted line.
  The original results are taken from Fig.~\ref{fig:dist1} with a red solid line.
  }\label{fig:PW}
\end{figure}

In Fig.~\ref{fig:PW}, we further calculate the dipole strength function with complex scaling,
where four valence neutrons in the $1^-$ states of $^8$He are treated as plane waves, namely, no final-state interaction is considered.
We adopt this condition by omitting $v^{\alpha N}$ and $v^{NN}$ in the Hamiltonian in Eq. (\ref{eq:Ham}) for the $1^-$ states.
This calculation is useful to understand the effect of the final-state interaction on the strength function
and we have performed a similar analysis for two-neutron halo nuclei of $^6$He and $^{11}$Li \cite{myo01,myo07b,myo08}.
It is found that the magnitude of strength becomes lower and the distribution is much wider over the excitation energy
in comparison with the original calculation including the interactions.
We confirm a mild bump at a higher excitation energy of around 40 MeV.
This indicates that the interaction effect in the $1^-$ states of $^8$He is important to explain the observed strength below 20 MeV shown in Fig. \ref{fig:dist1}.
It is known that the dipole strength to the plane waves reflects the spatial distribution of the weak-binding neutrons in the ground state \cite{nakamura94,myo98},
but this effect cannot explain the observed strength.
When we omit only the interactions between valence neutrons $v^{NN}$ in the Hamiltonian,
the strength starts to show at around 5 MeV near the $^5$He+$n$+$n$+$n$ threshold energy,
where $^5$He is a $3/2^-$ resonance,
and the peak of the strength shifts to around 10 MeV and the height is reduced from the original one.

\begin{figure}[t]
\centering
\includegraphics[width=8.5cm,clip]{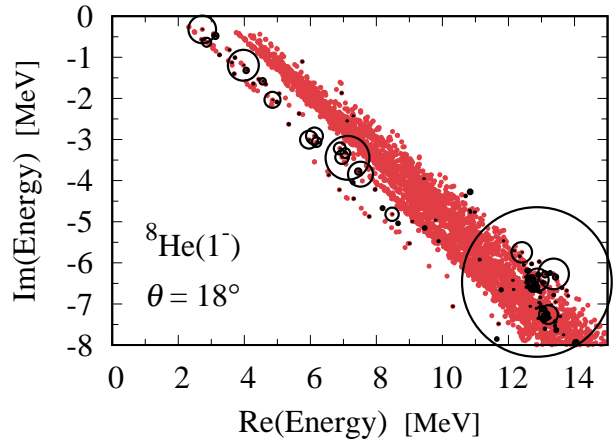}
\caption{
  Dipole matrix elements of the $1^-$ eigenstates of $^8$He using the complex scaling with $\theta=18^\circ$ in the complex energy plane, taken from Ref. \cite{myo22}. 
  Small red dots indicate the complex eigenenergies.
  The radius of the open circles is the absolute value of the real part of the dipole matrix element for each state in the arbitrary units.}
\label{fig:size}
\end{figure}

To clarify the structures of the original dipole strength,
in Fig.~\ref{fig:size}, we show the distribution of the dipole matrix elements for the $1^-$ eigenstates with the complex energy eigenvalues shown in Fig. \ref{fig:CSM},
which is originally explained in the previous letter \cite{myo22} and complementarily useful in the present analysis.
The size of the open circles is proportional to the real part of the following reduced dipole matrix element,
${\rm Re}\left\{\bras{\wtil{\Psi}^\theta_{0^+}}|\hO_{E1}^{\dagger,\theta}|\kets{\Psi^\theta_{1^-,\nu}}\bras{\wtil{\Psi}^\theta_{1^-,\nu}}|\hO^\theta_{E1}|\kets{\Psi^\theta_{0^+}}\right\}$.
Below 10 MeV of the excitation energy, the two-body continuum states of the $^7$He($3/2^-$)+$n$ channel show a large contribution to the dipole strength.
The strengths from the other channels of $^4$He+$n$+$n$+$n$+$n$, $^5$He+$n$+$n$+$n$, and $^6$He+$n$+$n$ are very minor.

\begin{table}[t]
  \begin{minipage}[t]{0.47\textwidth}
    \caption{Dominant parts of the squared amplitudes $(C_{c,\theta}^J)^2$ of the $1^-$ state of $^8$He with the $^7$He+$n$($s$-wave) configuration.}
    \label{tab:cmp1}
    \centering
    \begin{tabular}{c|ccc}
      \noalign{\hrule height 0.5pt}
      Configuration                            &  $(C_{c,\theta}^J)^2$ \\ \hline
      $(p_{3/2})^3(1s_{1/2})$                 &  $0.918-0.020i$       \\ 
      $(p_{3/2})(p_{1/2})^2(1s_{1/2})$        &  $0.027+0.001i$       \\ 
      $(p_{3/2})(d_{5/2})^2(1s_{1/2})$        &  $0.019+0.002i$       \\ 
      $(p_{3/2})^2(p_{1/2})(1s_{1/2})$        &  $0.014+0.007i$       \\ 
      \noalign{\hrule height 0.5pt}
    \end{tabular}
  \end{minipage}
  \hspace*{0.6cm}
  \begin{minipage}[t]{0.47\textwidth}
    \caption{Dominant parts of the squared amplitudes $(C_{c,\theta}^J)^2$ of the $1^-$ state of $^8$He with the $^7$He+$n$($d$-wave) configuration.}
    \label{tab:cmp2}
    \centering
    \begin{tabular}{c|ccc}
      \noalign{\hrule height 0.5pt}
      Configuration                            &  $(C_{c,\theta}^J)^2$ \\ \hline
      $(p_{3/2})^3(d_{5/2})$                  &  $0.918-0.013i$       \\ 
      $(p_{3/2})(p_{1/2})^2(d_{5/2})$         &  $0.028+0.005i$       \\ 
      $(p_{3/2})(d_{5/2})^3$                  &  $0.017-0.003i$       \\ 
      $(p_{3/2})(1s_{1/2})^2(d_{5/2})$        &  $0.006+0.002i$       \\ 
      $(p_{3/2})(d_{3/2})^2(d_{5/2})$         &  $0.004+0.001i$       \\ 
      \noalign{\hrule height 0.5pt}
    \end{tabular}
  \end{minipage}
\end{table}

In Tables \ref{tab:cmp1} and  \ref{tab:cmp2}, we select the typical $^7$He+$n$ states having relatively large dipole matrix elements
and show the dominant configurations of valence neutrons in these states; 
one is $^7$He+$n$($s$-wave) and the other is $^7$He+$n$($d$-wave).
The different radial components in the same orbit are summed up.
It is found that the $^7$He+$n$ channels dominantly have the $(p_{3/2})^3$ configuration for the $^7$He part,
and the last neutron is in the continuum states of $s$-wave or $d$-wave.
Hence, the dipole strengths to the $^7$He$+n$ channels shown in Fig.~\ref{fig:size}
are understood to be caused by the single particle excitation from the ground state of $^8$He.

In Fig.~\ref{fig:size}, several $1^-$ states show the large dipole matrix elements at around 13 MeV of the excitation energy.
In the previous letter \cite{myo22}, it is found that these states have a property of the strong configuration mixing
with the multineutron excitations.
This feature suggests the collectivity of four-neutron ($4n$) excitations and
we assign these states as the soft dipole mode of $4n$ oscillating against the $^4$He core.
The excitation energy of 13 MeV looks high but can be naively understood from the $1\hbar\omega$ excitation of the relative motion
between $4n$ and the $^4$He core estimated using the mean relative distance in the ground state \cite{myo22}.

\begin{figure}[t]
\centering
\includegraphics[width=8.5cm,clip]{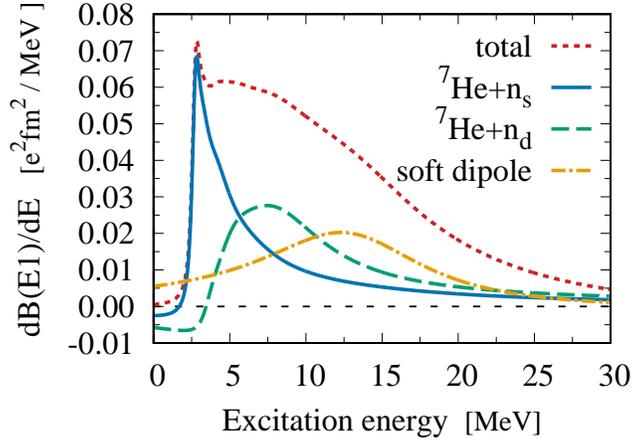}
\caption{
  Decomposition of the electric dipole strength of $^8$He using the complex scaling as a function of the excitation energy $E$,
  taken from Ref. \cite{myo22}. 
  The components are
  $^7$He(3/2$^-$)+$n$($s$-wave) with a blue solid line,
  $^7$He(3/2$^-$)+$n$($d$-wave) with a green dashed line,
  the soft dipole mode with an orange dash-dotted line,
  in addition to the total strength with a red dotted line.}
\label{fig:cmp}
\end{figure}

In Fig.~\ref{fig:cmp}, we show the results of the decomposition of the dipole strength
into the $^7$He+$n$ channels and the soft dipole mode by using the results of Figs. \ref{fig:CSM} and \ref{fig:size}.
It is confirmed that the $^7$He(3/2$^-$)+$n$ channel is dominant below 10 MeV of the excitation energy,
indicating the sequential breakup process via the $^7$He resonance.
At 13 MeV, the soft dipole mode makes a broad peak in the strength, which explains the total strength above 10 MeV as shown in Fig. \ref{fig:dist1}.

\subsection{Transition density}\label{sec:dns}

Transition density is useful to understand the spatial distribution of multineutron in the dipole excitations of $^8$He \cite{myo22}.
The transition density $\rho_{E1,\nu}(r)$ is defined to be the integrand of the reduced matrix element of the transition to the state $\nu$.
\begin{eqnarray}
  \bras{\wtil{\Psi}_{1^-,\nu}}|\hO_{E1}|\kets{\Psi_{0^+}}
  &=& \int_0^\infty \rho_{E1,\nu}(r)\, r^2 dr \, ,
  \\
\rho_{E1,\nu}(r)
  &=& -\frac{Z_\alpha}{A} e \ \sum_{j=1}^{A_v} \bras{\wtil{\Psi}_{1^-,\nu}}| r_j Y_1(\hat{\vc{r}}_j) \dfrac{\delta(r_j-r)}{r_j^2} |\kets{\Psi_{0^+}}  ,
\end{eqnarray}
where $r$ stands for the distance between $^4$He and a neutron.
It is possible to apply the complex scaling to this quantity, however, the complex scaling changes the spatial distribution of the integrand along the scaled coordinate of $r e^{i\theta}$.
To discuss the spatial property of the transition density $\rho_{E1,\nu}(r)$ directly,
we calculate the transition density with $\theta=0^\circ$ and select the specific states which have the same properties of the configurations as obtained with complex scaling. 

\begin{figure}[b]
\centering
\includegraphics[width=8.5cm,clip]{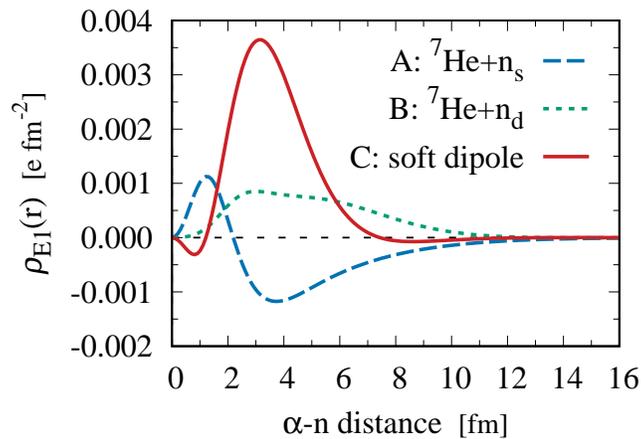}
\caption{
  Transition densities of electric dipole strengths for three $1^-$ states of $^8$He as shown in Fig. \ref{fig:disc1} with labels in units of $e$fm$^{-2}$.
  Three states are shown with their representative configurations of A: $^7$He+$n$($s$-wave) with a blue dashed line, B: $^7$He+$n$($d$-wave) with a green dotted line, and C: soft dipole mode with a red solid line, respectively.}
\label{fig:dns}
\end{figure}

We choose a typical state having the $^7$He+$n$ ($s$-wave) channel with a large dipole matrix element of $^8$He, which is the state with label A in Fig. \ref{fig:disc1}.
In this state, a single neutron is mainly excited from the $p$-state to the $s$-state, and the transition density is shown with the blue dashed line in Fig.~\ref{fig:dns}. 
The long tail behaviour is obtained in the distribution because of the separation of one neutron from $^8$He.
The second case is the $^7$He+$n$ ($d$-wave) channel in the state with label B in Fig. \ref{fig:disc1}.  
In Fig.~\ref{fig:dns}, this state also shows the long tail behaviour in the green dotted line, indicating the $d$-wave neutron in the continuum.
The last case is the soft dipole mode with label C in Fig. \ref{fig:disc1}.  
The transition density in the red solid line shows the large strength concentrating at 3 fm without a long tail,
indicating a soft dipole mode of $4n$ oscillating against the $^4$He core \cite{ikeda92}.

In the summary of the dipole transition densities of $^8$He,
there are two kinds of different excitations; the single-particle excitations to the $^7$He+$n$ channel
and the collective excitation to the soft dipole mode.

\subsection{Possibility of soft dipole resonance}

We locate an isolated $1^-$ resonance that represents the soft dipole mode of $^8$He.
For this purpose, we adopt a larger scaling angle $\theta$ than $18^\circ$ used in Fig. \ref{fig:CSM} in the complex scaling
and try to search for the resonance pole, which is separated from the rotated continuum states in the complex energy plane.
In the calculation of the complex scaling with a large $\theta$, the numerical accuracy generally decreases
because we expand the radial wave function with the finite number of the basis functions.
Due to this reason, several continuum lines in the complex energy plane often deviate from the position at the threshold energies.
It is possible to improve the numerical stability with a large $\theta$
in terms of the complex-range Gaussian basis functions \cite{hiyama03}, which increases the basis number.

\begin{figure}[b]
\centering
\includegraphics[width=8.5cm,clip]{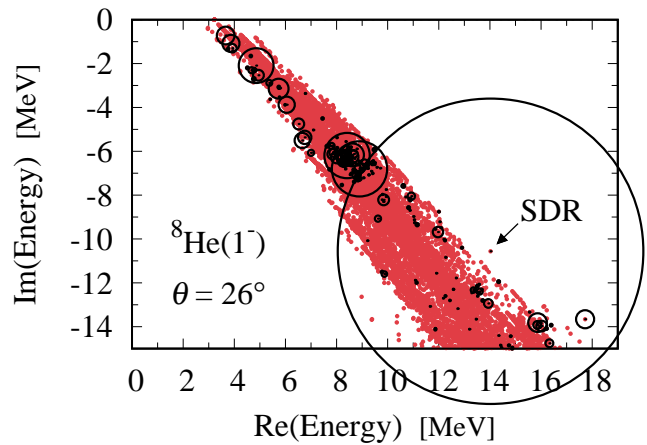}
\caption{
  Dipole matrix elements of the $1^-$ eigenstates of $^8$He using the complex scaling with $\theta=26^\circ$ in the complex energy plane measured from the ground state energy.
  Notations are the same as those used in Fig.~\ref{fig:size}.
  The short arrow indicates the eigenenergy of the soft dipole resonance (SDR) with $(E_x,\Gamma)=(14, 21)$ MeV showing the large dipole matrix element.
}
\label{fig:size2}
\end{figure}

In Fig. \ref{fig:size2}, we show the results of the distribution of the dipole matrix elements using $\theta=26^\circ$ in the complex energy plane,
together with the complex energy eigenvalues, similarly to Fig. \ref{fig:size}.
We obtain one candidate of the resonance pole at $(E_x,\Gamma)=(14,21)$ MeV having a large dipole matrix element shown by the large size of the circle.
Here, large $\theta$ values decrease numerical accuracy, so the low-energy continuum lines such as the $^7$He+$n$ channel are slightly shifted to the right direction and not distinguishable clearly.
On the other hand, the resonance pole is stably obtained with this value of $\theta$ in the complex energy plane.
We consider that this resonance becomes the candidate for the soft dipole resonance (SDR) of $^8$He.

We evaluate the radial properties of the SDR in Table \ref{tab:radius2}.
For unbound states with complex energies, the matrix elements can be complex, and if the state is a resonance, its radius is uniquely determined \cite{myo20,burgers96,dote18}.
We can interpret the spatial size of the resonance using the real part of the radius if the imaginary part is relatively smaller than the real part.

It is found that the obtained radial matrix elements of the SDR have larger real parts than the imaginary parts, although this resonance has a very large decay width of 21 MeV.
Using real parts, the resonance shows the large spatial sizes in comparison with those of the ground state as shown in Table \ref{tab:radius}.
The radius of $4n$ in the resonance is 3.71 fm,  which is larger than 2.91 fm in the ground state. 
The distance between $^4$He and $4n$ also becomes large as 2.67 fm from 2.05 fm in the ground state.
These results indicate that in the SDR,
the size of $4n$ is expanded from that in the ground state and the mean relative distance also becomes large for dipole oscillation.

\begin{table}[t]  
  \caption{Radial properties of the soft dipole resonance of $^8$He in units of fm.
    We show matter, proton, and neutron radii, and mean distance between $^4$He and four valence neutrons $\alpha\mbox{-}4n$,
    and the radius of four valence neutrons $4n$.}
  \label{tab:radius2}
  \centering
    \begin{tabular}{c|ccc}
      \noalign{\hrule height 0.5pt}
      matter         & $3.11+0.86i$  \\
      proton         & $1.97+0.28i$  \\
    ~~neutron~~      & $3.41+0.99i$  \\
$\alpha\mbox{-}4n$   & $2.67+0.84i$  \\
$4n$                 & $3.71+1.14i$  \\
      \noalign{\hrule height 0.5pt}
\end{tabular}
\end{table}

\begin{figure}[t]
\centering
\includegraphics[width=8.5cm,clip]{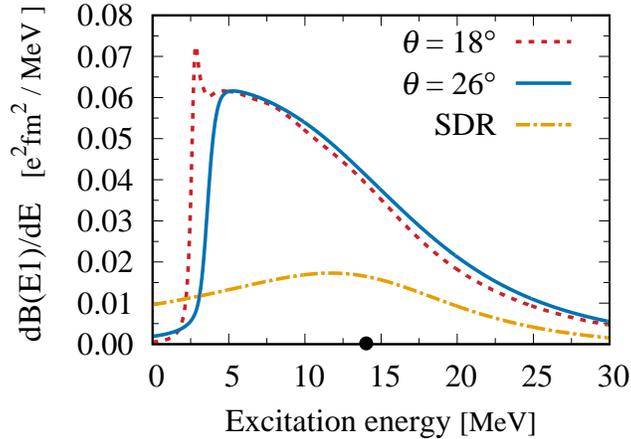} 
\caption{
  Electric dipole strength $dB(E1)/dE$ of $^8$He as a function of the excitation energy using the complex scaling with $\theta=26^\circ$ (blue solid line) and $\theta=18^\circ$ (red dotted line).
  Units are in $e^2$fm$^2$/MeV.
  The contribution of soft dipole resonance (SDR) with $\theta=26^\circ$ is given with an orange dash-dotted line.
  The resonance energy is shown by the black solid circle on the energy axis.}
\label{fig:cmp2}
\end{figure}

Finally, we calculate the electric dipole strength of $^8$He using the complex-scaled solutions including the SDR obtained with $\theta=26^\circ$ in Fig. \ref{fig:cmp2}.
We also compare the results with those obtained using $\theta=18^\circ$.
It is found that strength at around 3 MeV of the excitation energy disappears.
because the continuum states of the $^7$He+$n$ channel are slightly shifted in the complex energy plane as shown in Fig. \ref{fig:size2}.
Above 5 MeV of the excitation energy, the strength is almost common between two results of $\theta=26^\circ$ and $18^\circ$.
We further extract the component of the SDR alone, which is described following Eq.~(\ref{eq:strength3}) as 
\begin{eqnarray} 
\frac{dB(E1,{\rm SDR})}{dE}&=&-\frac{1}{\pi}{\rm Im}\left\{\frac{M_{\rm R}+iM_{\rm I}}{E-E_{\rm SDR}}\right\}
~=~\frac{1}{\pi}\frac{M_{\rm R}\Gamma/2-M_{\rm I}(E-E_x)}{(E-E_x)^2+\Gamma^2/4},
\label{eq:SDR}
\end{eqnarray}
where 
\begin{equation}
E_{\rm SDR}~=~E_x-i\Gamma/2, \qquad
\langle \tilde{\Psi}^\theta_{0^+}||(O^\dagger_{E1})^\theta||\Psi^\theta_{\rm SDR}\rangle
\langle \tilde{\Psi}^\theta_{\rm SDR}||O^\theta_{E1}||\Psi^\theta_{0^+}\rangle~=~M_{\rm R}+iM_{\rm I}.
\end{equation}
The obtained distribution of SDR in Fig. \ref{fig:cmp2} is similar to those of the soft dipole mode as shown in Fig. \ref{fig:cmp}.
This indicates that the SDR is the main component of soft dipole mode in the dipole strength.
The integration of Eq.~(\ref{eq:SDR}) over energy becomes $M_{\rm R}$.

The SDR provides the electric dipole matrix element of $M_{\rm R}+iM_{\rm I}=0.55+0.25i$ ($e^2$fm$^2$)
and the real part $M_{\rm R}$ exhausts about half of the cluster sum-rule value $B_{\rm c}(E1)$=1.01 $e^2$fm$^2$ presented in Table \ref{tab:radius}.
This property represents the strong collectivity of the $4n$ excitation in the SDR from the ground state of $^8$He.
The imaginary part $M_{\rm I}$ determines the deviation of the energy dependence of the strength function from the Breit-Wigner distribution
according to Eq.~(\ref{eq:SDR}) \cite{morimatsu94},
and the peak energy of the strength is shifted from the resonance energy $E_x$ as shown by the solid circle in Fig. \ref{fig:cmp2}.

\section{Summary}\label{sec:summary}

We investigated the possibility of the soft dipole resonance of $^8$He
and its effect on the electric dipole strength using the $^4$He+$n$+$n$+$n$+$n$ five-body cluster model.
We describe the multineutron motion around the $^4$He core in the cluster orbital shell model (COSM) and the complex scaling method (CSM)
and obtain many-body resonant and non-resonant continuum states of $^8$He.
The Green's function is utilized with the complex scaling to calculate the dipole strength function of $^8$He.

It is found that the dipole strength of $^8$He shows the low-energy enhancement and the distribution continues in a wide energy range up to around 20 MeV.
Using the extended completeness relation (ECR) expanded in the complex-scaled eigenstates of $^8$He, the strength is mainly constructed by two components.
Below 10 MeV of the excitation energy, two-body continuum states consisting of the resonance of $^7$He and a neutron contribute to the strength.
This is the single-particle excitations and indicates the sequential breakup process via $^7$He. 
At around 13 MeV, the collective excitations of four neutrons ($4n$) make a broad peak in the strength.
From the analysis of the transition densities of the dipole matrix elements,
this excitation causes the soft dipole mode, in which $4n$ are oscillating against the inert $^4$He core \cite{ikeda92},
and is different from the giant dipole resonance, in which the core nucleus is excited.

For the soft dipole mode, we searched for the resonance pole in the complex energy plane by using a large $\theta$ in the complex scaling.
As a result, we found one candidate of the soft dipole resonance at the excitation energy of 14 MeV with the decay width of 21 MeV,
which largely contributes to the dipole strength in the energy region of 13 MeV.
These interesting features of the dipole excitations of $^8$He would be confirmed by experiments, such as the Coulomb breakup reaction of $^8$He \cite{lehr22}. 

\section*{Acknowledgments}
This work was supported by JSPS KAKENHI Grants No. JP18K03660, No. JP20K03962, and No. JP22K03643.
Numerical calculations were partly achieved through the use of large-scale computer systems, SQUID, at the Cybermedia Center, Osaka University.

\def\JL#1#2#3#4{ {{\rm #1}} \textbf{#2}, #3 (#4)}  
\nc{\PPNP}[3]   {\JL{Prog. Part. Nucl. Phys.}{#1}{#2}{#3}.}
\nc{\PTEP}[3]   {\JL{Prog. Theor. Exp. Phys.}{#1}{#2}{#3}.}
\nc{\PRep}[3]   {\JL{Phys. Rep.}{#1}{#2}{#3}.}


\begin{thebibliography}{00}

\bibitem{tanihata85} I.~Tanihata~{\it et al.},~\PRL{55,2676,1985} 
\bibitem{tanihata13} I. Tanihata, H. Savajols, R. Kanungo, \PPNP{68}{215}{2013}  

\bibitem{hansen87} P.G. Hansen and B.Jonson, \JL{Europhys. Lett.}{4}{409}{1987}. 
\bibitem{ikeda92}  K. Ikeda, \NPA{538,355c,1992}  

\bibitem{nakamura94} T.~Nakamura {\it et al.}, \PLB{331,296,1994} 
\bibitem{aumann99} T. Aumann {\it et al.}, \PRC{59,1252,1999}  
\bibitem{nakamura06} T.~Nakamura {\it et al.}, \PRL{96,252502,2006} 

\bibitem{tilley04}    D. R. Tilley, J. H. Kelley, J. L. Godwin, D. J. Millener, J. E. Purcell, C. G. Sheu, and H.R. Weller, \NPA{745,155,2004} 
\bibitem{korsheninnikov93} A.A. Korsheninnikov {\it et al.}, \PLB{316,38,1993} 
\bibitem{iwata00}    Y. Iwata {\it et al.}, \PRC{62,064311,2000}  
\bibitem{meister02}  M. Meister {\it et al.}, \NPA{700,3,2002} 
\bibitem{chulkov05}  L.V. Chulkov {\it et al.}, \NPA{759,43,2005} 
\bibitem{mueller07}  P. Mueller {\it et al.}, \PRL{99,252501,2007} 
\bibitem{golovkov09} M.S. Golovkov {\it et al.},~\PLB{672,22,2009} 
\bibitem{holl21}   M. Holl {\it et al.}, \PLB{822,136710,2021} 

\bibitem{nonaiti22}  F. Bonaiti, S. Bacca, G. Hagen, \PRC{105,034313,2022}
\bibitem{piekarewicz22}  J. Piekarewicz, \PRC{105,044310,2022}

\bibitem{myo10}      T. Myo, R. Ando, and K. Kat\=o, \PLB{691,150,2010} 
\bibitem{myo12}      T. Myo, Y. Kikuchi, and K. Kat\=o, \PRC{85,034338,2012} 
\bibitem{myo14b}     T. Myo and K. Kat\=o, \PTEP{2014}{083D01}{2014} 
\bibitem{myo21}      T. Myo, M. Odsuren, K. Kat\=o, \PRC{104,044306,2021}  
\bibitem{myo22}      T. Myo, K. Kat\=o, \PRC{106,L021302,2022} 

\bibitem{suzuki88}   Y. Suzuki and K. Ikeda, \PRC{38,410,1988} 
\bibitem{masui06}    H. Masui, K. Kat\=o,~K. Ikeda, \PRC{73,034318,2006}
\bibitem{masui12} H. Masui, K. Kat\=o, K. Ikeda \NPA{895,1,2012} 
\bibitem{myo07a}     T. Myo, K. Kat\=o, and K. Ikeda,~\PRC{76,054309,2007} 

\bibitem{ho83}       Y. K. Ho, \PRep{99}{1}{1983} 
\bibitem{moiseyev98} N. Moiseyev, \PRep{302}{211}{1998} 
\bibitem{aoyama06}   S. Aoyama, T. Myo, K. Kat\=o, K. Ikeda, \PTP{116,1,2006} 
\bibitem{moiseyev11} N. Moiseyev, {\it Non-Hermitian quantum mechanics} (Cambridge University Press, Cambridge, 2011). 
\bibitem{myo14a}     T. Myo, Y. Kikuchi, H. Masui, K. Kat\=o, \PPNP{79}{1}{2014}  
\bibitem{myo20}      T. Myo, K. Kat\=o, \PTEP{2020}{12A101}{2020}  

\bibitem{myo98}      T. Myo, A.~Ohnishi, and K. Kat\=o, \PTP{99,801,1998} 
\bibitem{myo01}      T. Myo, K. Kat\=o, S. Aoyama and K. Ikeda, \PRC{63,054313,2001} 
\bibitem{suzuki05}   R. Suzuki, T.~Myo and K. Kat\=o, \PTP{113,1273,2005} 
\bibitem{odsuren15}  M. Odsuren, Y. Kikuchi, T. Myo, M. Aikawa, and K. Kat\=o, \PRC{92,014322,2015} 

\bibitem{kikuchi10}  Y. Kikuchi, K. Kat\=o, T. Myo, M. Takashina, K. Ikeda, \PRC{81,044308,2010}  
\bibitem{kikuchi13}  Y. Kikuchi, T. Myo, K. Kat\=o, K. Ikeda, \PRC{87,034606,2013}  

\bibitem{lehr22}  C. Lehr et al. (SAMURAI Collaboration) (unpublished).  

\bibitem{kanada79}   H.~Kanada,~T.~Kaneko,~S.~Nagata,~M.~Nomoto,~\PTP{61,1327,1979}  
\bibitem{tang78}     Y. C. Tang, M. LeMere and D. R. Thompson, \PRep{47}{167}{1978}  
\bibitem{saito69}    S. Saito \PTP{41,705,1969}  

\bibitem{hiyama03} E. Hiyama, Y. Kino, M. Kamimura, \PPNP{51}{223}{2003} 

\bibitem{ABC}        J. Aguilar and J.M. Combes, \JL{Commun. Math. Phys.}{22}{269}{1971}, 
	             E. Balslev and J.M. Combes, \JL{Commun. Math. Phys.}{22}{280}{1971}. 

\bibitem{berggren68} T. Berggren,~\NP{A109,265,1968} 
\bibitem{kruppa07} A. T. Kruppa, R. Suzuki and K. Kato, \PRC{75,044602,2007} 
\bibitem{dote13}   A. Dot\'e, T. Inoue, T. Myo, \NPA{912,66,2013} 

\bibitem{odsuren14}  M. Odsuren, K. Kat\=o, M. Aikawa and T. Myo, \PRC{89,034322,2014} 
\bibitem{odsuren21}  M. Odsuren, T. Myo, Y. Kikuchi, M. Teshigawara and K. Kat\=o, \PRC{104,014325,2021} 

\bibitem{tanihata92} I.~Tanihata~{\it et al.},~\PLB{289,261,1992}  
\bibitem{alkazov97}  G. D. Alkhazov {\it et al.}, \PRL{78,2313,1997} 
\bibitem{kiselev05}  O. A. Kiselev {\it et al.}, \JL{Eur. Phys. J. A25}{Suppl.1}{215}{2005}. 
\bibitem{brodeur12}  M. Brodeur {\it et al.}, \PRL{108,052504,2012} 
\bibitem{krauth21}   J.J. Krauth {\it et al.}, \JL{Nature}{589}{527}{2021}. 

\bibitem{myo07b}     T. Myo, K. Kat\=o, H. Toki, and K. Ikeda, \PRC{76,024305,2007} 
\bibitem{myo08}      T. Myo, Y. Kikuchi, K. Kat\=o, H. Toki, and K. Ikeda, \PTP{119,561,2008} 

\bibitem{burgers96}   A. B\"urgers and J.M. Rost, \JL{J. Phys. B}{29}{3825}{1996}. 
\bibitem{dote18}   A. Dot\'e, T. Inoue, T. Myo, \PLB{784,405,2018} 
\bibitem{morimatsu94}  O. Morimatsu, K. Yazaki, \PPNP{33}{679}{1994}  

\end{thebibliography}
\end{document}